\documentclass[prl,aps,twocolumn,showpacs,amsmath,amssymb]{revtex4}
\usepackage{graphicx}% Include figure files
\usepackage{dcolumn}% Align table columns on decimal point
\usepackage{bm}% bold math

\begin{document}
\title{Superconductivity at 56 K in Samarium-doped SrFeAsF}
\author{G. Wu, Y. L. Xie, H. Chen, M. Zhong, R. H. Liu, B. C. Shi,
Q. J. Li, X. F. Wang, T. Wu, Y. J. Yan, J. J. Ying}
\author{X. H. Chen}
\altaffiliation{Corresponding author} \email{chenxh@ustc.edu.cn}
\affiliation{Hefei National Laboratory for Physical Science at
Microscale and Department of Physics, University of Science and
Technology of China, Hefei, Anhui 230026, China\\}
\date{\today}

\begin{abstract}

We synthesized the samples Sr$_{1-x}$Sm$_x$FFeAs with
ZrCuSiAs-type structure. These samples were characterized by
resistivity and susceptibility. It is found that substitution of
rare earth metal for alkaline earth metal in this system
suppresses the anomaly in resistivity and induces
superconductivity. Superconductivity at 56 K in nominal
composition Sr$_{0.5}$Sm$_{0.5}$FFeAs is realized, indicating that
the superconducting transition temperatures in the iron arsenide
fluorides can reach as high as that in oxypnictides with the same
structure.
\end{abstract}


\maketitle
\newpage

The recent discovery of superconductivity in oxypnictides with the
critical temperature ($T_C$) higher than McMillan limit of 39 K
(the theoretical maximum predicted by BCS theory) has generated
great excitement\cite{yoichi,chen,chen1,ren,liu,wang} since the
superconductivity is clearly unconventional compared with in the
cuprate superconductors. The high-$T_c$ iron pnictides with
ZrCuSiAs-type structure adopt a layered structure of alternating
(FeAs)$^-$ and (LnO)$^+$ layers with eight atoms in a tetragonal
unit cell. Structural phase transition from tetragonal to
orthorhombic happens before the antiferromagnetic
spin-density-wave ordering.\cite{cruz,margadonna} Such transition
leads to an anomaly in resistivity. Doping of charge into the
system suppresses the structural and magnetic instabilities, and
induces superconductivity. Recently, the iron arsenide fluorides
AEFeAsF (AE=Sr, Ca etc.) with ZrCuSiAs-type structure, where the
(LnO)$^+$ layers in LnFeAsO are replaced by (AEF)$^+$ layer have
been reported.\cite{matsuishi,tegel,han} Co-doping in CaFeAsF
leads to a maximum $T_c$ of 22 K,\cite{matsuishi} and a
superconducting transition at about 36 K is reported in
Sr$_{0.8}$La$_{0.2}$FeAsF.\cite{zhu}

Here we report the discovery of superconductivity at 56 K in
nonminal composition Sr$_{1-x}$Sm$_x$FeAsF (x=0.5) with
ZrCuSiAs-type structure. The $T_c$ is almost the same as the
highest superconducting transition temperature observed in F-doped
oxypnictide superconductors.

The polycrystalline samples with nominal composition of
Sr$_{1-x}$Sm$_x$FeAsF were synthesized by solid state reaction
method by using SrF$_2$, SrAs, SmAs, and Fe$_2$As as starting
materials: 0.5SrF$_2$ ${+}$ (0.5-$x$)SrAs ${+}$ $x$SrAs ${+}$
0.5Fe$_2$As $\rightarrow$ Sr$_{1-x}$Sm$_x$FeAsF. SrAs was
pre-synthesized by heating the mixture of Sr powder and As powder
in an evacuated quartz tube at 873 K for 10 hours. SmAs and
Fe$_2$As were obtained by respectively reacting the mixture of Sm
powder, Fe powder and As powder in evacuated quartz tubes at 1073
K for 10 hours. The raw materials were accurately weighed
according to the stoichiometric ratios of Sr$_{1-x}$Sm$_x$FeAsF,
then the weighed powders were thoroughly grounded and pressed into
pellets. The pellets were wrapped with Ta foil and sealed in
evacuate quartz tubes. The SrFeAsF was slowly heated to 1173 K and
kept at this temperature for 40 hours and cooled down to room
temperature. Then the resultant pellet was grounded again, sealed
in a quartz tube for a second sintering at 1273 K for 20 hours.
The samples of Sr$_{0.8}$Sm$_{0.2}$FeAsF and
Sr$_{0.5}$Sm$_{0.5}$FeAsF were slowly heated to 1273 K for 20
hours, then the products were grounded again, sealed in a quartz
tube for a second sintering at 1273 K for 10 hours. The sample
preparation process except for annealing was carried out in glove
box (O$_2$, H$_2$O $<$ 1 ppm) in which high pure argon atmosphere
is filled.

The crystal structure of these samples was characterized by X-ray
diffraction (XRD) with Rigaku D/max-A X-ray diffractometer using
Cu K$_\alpha$ radiation ($\lambda$=1.5418\AA) in the 2$\theta$
range of 5 - 65$^{o}$ with the step of 0.02$^{o}$ at room
temperature. Resistivity measurements were performed on a AC
resistance bridge (Linear Research Inc., Model LR700) by the
standard four-probe method. The measurement of susceptibility was
performed in Quantum Design PPMS systems (Quantum Design).

Figure 1 shows X-ray powder diffraction patterns of SrFeAsF,
Sr$_{0.8}$Sm$_{0.2}$FeAsF and Sr$_{0.5}$Sm$_{0.5}$FeAsF
respectively. All diffraction peaks in the XRD pattern of the
sample SrFeAsF can be indexed by a tetragonal structure with
a=0.3995 nm and c=0.8961 nm, where no impurity peak is observed.
It indicates that the samples is single phase. For
Sr$_{0.8}$Sm$_{0.2}$FeAsF sample, all the main peaks can be
indexed to ZrCuSiAs-type structure with a=0.3929 nm and c=0.8958
nm. Only small amount of impurity phase of SrF$_2$ was detected.
For Sr$_{0.5}$Sm$_{0.5}$FeAsF sample, the diffraction peaks except
for the diffraction peaks from impurity phases of SrF$_2$ and SmAs
can be indexed by a tetragonal structure with a=0.3918 nm and
c=0.8956 nm. It is found that Sm doping leads to an apparent
decrease in a-axis and a slight contraction of the c-axis. With
increasing Sm doping, impurity phase of SrF$_2$ apparently
increases. The impurity phase of SmF shows up in
Sr$_{0.5}$Sm$_{0.5}$FeAsF sample.

\begin{figure}
\includegraphics[width=9 cm]{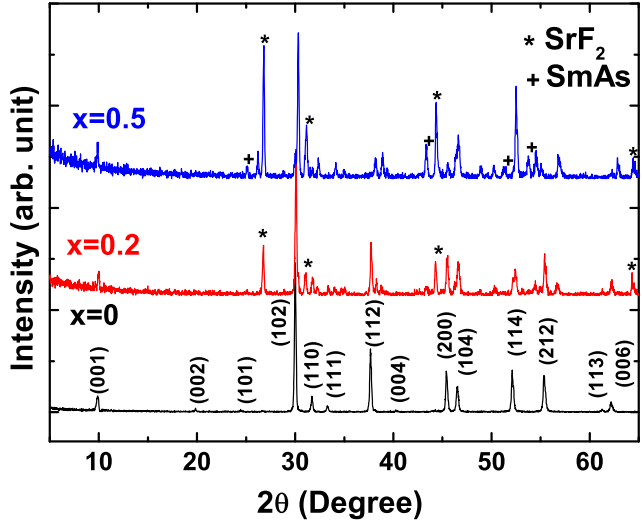}
\caption{ X-ray diffraction patterns at room temperature for the
samples Sr$_{1-x}$Sm$_x$FeAsF with x=0, 0.2, and 0.5. All
diffraction peaks can be indexed with a tetragonal stucture for
the x=0 sample. However, the impurity phases of $SrF_2$ and SmAs
are observed in x=0.2 and 0.5 samples.}
\end{figure}

Figure 2 shows temperature dependence of resistivity for the
samples SrFeAsF, Sr$_{0.8}$Sm$_{0.2}$FeAsF and
Sr$_{0.5}$Sm$_{0.5}$FeAsF. The resistivity of SrFeAsF shows a
clear anomaly in resistivity at about 173 K. While cooling down
from temperature, the sample's resistivity grows slightly before
173K, then drops sharply with the temperature continues
decreasing. At low temperatures, the resistivity shows a
semiconducting behavior. This is consistent with previous
reports.\cite{tegel, han} The anomaly in resistivity is ascribed
to the formation of an SDW order or a structural transition,
similar to what's observed in LnFeAsO (Ln = rare earth elements)
and MFe$_2$As$_2$ (M = Ba, Sr) system. With Sm doping, the anomaly
in resistivity is suppressed, and the transition temperature
down-shifts to 163 K for Sr$_{0.8}$Sm$_{0.2}$FeAsF, while the
semiconducting behavior at low temperature is also suppressed. As
shown in Fig.2, a sharp superconducting transition at about 56 K
occurs in resistivity for the sample Sr$_{0.5}$Sm$_{0.5}$FeAsF.
The resistivity shows a metallic behavior in the whole temperature
range and no anomaly appears. It suggests that the SDW order or
the structural transition is completely suppressed for the
superconducting sample. The behavior of resistivity in the normal
state is exactly the same as that observed in superconducting
sample SmFeAsO$_{1-x}$F$_x$ (x=0.15 and 0.2).\cite{liu} It should
also be emphasized that no superconducting transition is observed
in the sample Sr$_{0.8}$Sm$_{0.2}$FeAsF.

\begin{figure}
\includegraphics[width=9 cm]{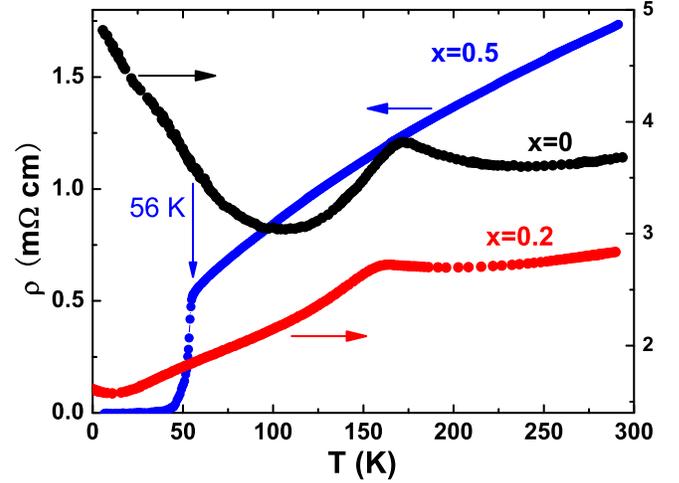}
\caption{ Temperature dependence of resistivity for the samples
Sr$_{1-x}$Sm$_x$FeAsF with x=0, 0.2, and 0.5. The anomaly in
resistivity associated with structural and SDW transitions is
suppressed with Sm doping. A sharp superconducting transition at
56 K is observed in the x=0.5 sample. }
\end{figure}

To confirm the superconductivity observed in resistivity for the
sample with nominal composition Sr$_{0.5}$Sm$_{0.5}$FeAsF, the
susceptibility is measured under 10 Oe in zero-field cooled and
field-cooled cycles. Temperature dependence of susceptibility is
shown in Fig.3. A clear diamagnetic transition occurs at 53.5 K
corresponding to the mid-transition temperature in resistivity,
indicating of a bulk superconductivity. It should be pointed out
that the superconducting volume is small, and the apparent
Meissner fraction is less than 10\%, while the resistance goes to
zero. The reason could be that there are some magnetic impurities
(such as FeAs) in this sample. The magnetic ordering of these
impurities may depress the Meissner fraction. The actual Meissner
fraction must be larger. Next step is how to synthesize the sample
of high purity and improve the superconducting volume fraction.
The superconducting phase should be Sm-doped SrFeAsF because only
impurity phases SrF$_2$ and SmAs are observed in x-ray diffraction
pattern of Sr$_{0.5}$Sm$_{0.5}$FeAsF. These impurity phases are
non-superconducting.

\begin{figure}
\includegraphics[width=9 cm]{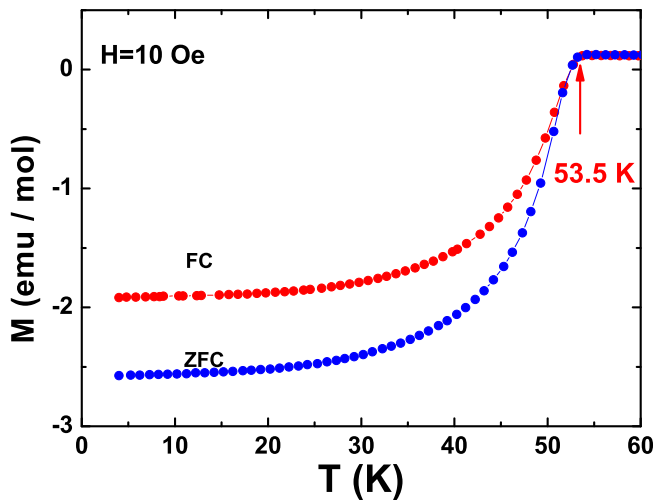}
\caption{Temperature dependence of DC magnetization for the sample
with nominal composition Sr$_{0.5}$Sm$_{0.5}$FeAsF measured at a
DC field of 10 Oe in the zero field-cooled (ZFC) and field-cooled
cycles. It indicates a bulk superconductivity.}
\end{figure}

In summary, partial substitution of Sr with Sm leads to the
suppression of the structural and SDW transition, meanwhile
induces superconductivity. The electrical conductivity and
magnetization measurements demonstrate a bulk superconductivity at
56 K in a nominal composition of Sr$_{0.5}$Sm$_{0.5}$FeAsF. Our
results indicate that it is possible to find superconductivity in
other fluorine-arsenide family.

 \vspace*{2mm} {\bf Acknowledgment:} This
work is supported by the Nature Science Foundation of China and by
the Ministry of Science and Technology of China (973 project No:
2006CB601001) and by National Basic Research Program of China
(2006CB922005).

\end{document}